\documentclass[12pt]{article}
\usepackage{geometry}
\usepackage{graphicx}
\usepackage{pdflscape}
\usepackage{multirow}
\usepackage{ragged2e}
\usepackage{rotating}
\usepackage{changepage} 
\usepackage{siunitx}
\usepackage{tcolorbox}
\usepackage{caption}
\usepackage{booktabs,calc}
\usepackage{array}
\usepackage{tabularx}
\usepackage{verbatim}
\usepackage{mathrsfs}
\usepackage{subfig}
\usepackage{setspace}
\usepackage{natbib}
\usepackage{amsmath}
\usepackage{amssymb}
\usepackage{amsthm}
\usepackage{enumerate}
\usepackage[linktocpage=true, colorlinks=true, linkcolor=blue, citecolor=black]{hyperref}

\usepackage{authblk}
\usepackage{multicol}
\usepackage{array}
\usepackage[section]{placeins}
\usepackage{threeparttable}
\usepackage[usenames, dvipsnames]{xcolor}
\newcolumntype{J}[1]{>{\justifying\arraybackslash}p{#1}}
\def\sym#1{\ifmmode^{#1}\else\(^{#1}\)\fi}
\newcolumntype{C}[1]{>{\centering\let\newline\\\arraybackslash\hspace{0pt}}m{#1}}
\usepackage[toc,page]{appendix} 

\usepackage{times}
\usepackage{mathptmx}

\begin{document}
\title{Gender and Agricultural Commercialization in Sub-Saharan Africa: Evidence from Three Panel Surveys}

\author[1]{Wei Li\thanks{Corresponding author \\
  \phantom{***} E-mail address: weili@tamu.edu (Wei Li)}} 
\author[1]{Kashi Kafle}
\author[2]{Anna Josephson}
\affil[1]{\small \emph{Texas A\&M University, College Station, Texas}}
\affil[2]{\small \emph{The University of Arizona, Tucson, Arizona}}

\date{} 
\maketitle

\begin{abstract}
\setlength{\baselineskip}{0.75\baselineskip} 

\noindent Agricultural commercialization is promoted as a key driver of development in Sub-Saharan Africa, yet its benefits may not extend equally to all farmers. Using longitudinal household data (LSMS-ISA) and a two-way Mundlak estimator, we investigate the relationship between farmers’ gender and agricultural commercialization in Ethiopia, Nigeria, and Tanzania. In Ethiopia and Nigeria, women-headed households and those with a higher share of women-managed land area face substantial disadvantages in market engagement, which are magnified in households oriented towards self-consumption. In both Ethiopia and Nigeria, women-headed households that sell crops are actually more likely to sell to market buyers and less likely to sell to individual buyers compared to men-headed households. Conversely, in Tanzania, the negative associations between gender and commercialization are weaker and less robust across all outcomes. Overall, these findings demonstrate that gender gaps in commercialization are highly context-specific rather than universal, highlighting the need for country-tailored policies that address the institutional and market constraints faced by women farmers.

\vspace{0.1in}
\end{abstract}

\newpage
\section{Introduction}

Agricultural commercialization in Sub-Saharan Africa (SSA) is promoted as a pathway to higher incomes, poverty reduction, and improved food security \citep{giller_food_2020,hilson_farming_2016}. Yet, the benefits of commercialization may not be equitably shared. Prior research documents systematic gender differences in access to land, inputs, credit, extension, and market networks, with implications for who participates and who gains in the process of commercialization \citep{doss_mens_2002, kilic_caught_2015, palacios2017, aguilar2015decomposition, quisumbing_womens_2021, josephson2025intra}. These constraints, often rooted in social norms and market frictions, can limit women's opportunities in output markets and reduce both the inclusiveness and efficiency of commercialization.  

This raises the central question we explore in this paper: how does the gender of the farmer influence the process and outcomes of agricultural commercialization in SSA? We study Ethiopia, Nigeria, and Tanzania using three waves of the World Bank’s Living Standard Measurement Study - Integrated Survey in Agriculture (LSMS-ISA). We examine two household-level gender measures: women’s headship and the share of land area managed by women. We then relate them to three indicators of commercialization: (1) participation in crop sales, (2) the crop commercialization index (CCI),\footnote{It is defined as the ratio of the gross value of crop sales to the gross value of crop harvest \citep{carletto_agricultural_2017}.} and (3) the cash-crop sales ratio. Because decisions regarding crop use may be made jointly, we also consider an alternative proxy for agricultural decision-making: the gender of the individual with authority over harvest allocation \citep{doss_intrahousehold_2013, anderson_husband_2017, josephson2025intra}. To explore mechanisms, we further disaggregate sales by buyer type, market type, and location, and examine the role of self-consumption in mediating market participation.

Causal identification is complicated by non-random selection into headship and management as well as by time-varying shocks that may jointly influence gender roles and market outcomes. Therefore, we estimate associations using the two-way Mundlak (TWM) estimator, which yields the same ``within'' estimates as two-way fixed effects (TWFE) for time-varying regressors while permitting coefficients on time-invariant variables, such as the gender of the household head, by augmenting a random-effects specification with household- and period-specific means of time-varying covariates \citep{wooldridge_two-way_2021}. We interpret all coefficients as associations rather than causal effects.

Three results emerge. First, in Ethiopia and Nigeria, and to a lesser degree in Tanzania, women-headed households and those with a larger share of women-managed land area are less likely to participate in crop sales, have lower CCI, and devote smaller sales shares to cash crops. Second, contrary to the narrative that women avoid or are prohibited from markets, we find that, conditional on selling, women are at least as likely to use market outlets as men. We find that in Nigeria, women’s sales are more proximate and informal, pointing to mobility and intermediation frictions, such as limited access to informal markets, exclusion from networks, rather than lower market orientation. Third, self-consumption is strongly negatively associated with market participation in Ethiopia and Nigeria, with larger penalties where women head households or manage more land. This helps us to reconcile lower overall sales with comparable use of market channels. Pooled multi-country estimates confirm these patterns, and a decision-maker proxy based on authority over harvest allocation is not statistically associated with commercialization when accounting for observable traits.

Our work complements much of the existing literature. The current body of literature shows that gender shapes farmers' participation in and benefits from agricultural commercialization \citep{doss_mens_2002,hill_mainstreaming_2014,maereka_womens_2023}. A large body of evidence documents persistent productivity gaps between men and women farmers, not driven by skill differences but rather by unequal access to land, technology, inputs, and extension services \citep{doss_designing_2001,quisumbing_womens_2021,fischer_gender_2012,meinzen-dick_womens_2019}. Intra-household studies suggest that commercialization can shift control over assets and income from women to men \citep{quisumbing_resources_2003}. Crop choice may also have gendered dynamics. \cite{doss_mens_2002} notes that while some crops are identified as men's crops, none are identified as women's. Commercialization is further shaped by factors such as market distance, road quality, and transportation costs. Burdens in these domains further restrict women's participation in markets \citep{woldie_contribution_2011,tilahun_factors_2023, oduol_womens_2017,kafle_reducing_2022}. 

With this paper, we make three contributions to the literature. First, we are among the first to examine the relationship between farmers' gender and agricultural commercialization in SSA, with a focus on output markets. This complements literature that has largely emphasized gender differences in productivity \citep{udry_gender_1996,aguilar2015decomposition,kilic_caught_2015}, resource access \citep{goldstein_profits_2008,peterman_review_2014,fischer_gender_2012}, and household welfare \citep{quisumbing_resources_2003,doss_effects_2006}. Next, we use a multi-year, nationally representative panel dataset from three countries, which allows us to observe trends and changes over time. This is in contrast to studies that are regional, within a single year, and/or qualitative \citep{pelekamoyo_access_2019,gupta_womens_2017,baden_womens_2013, msosa_challenges_2022}. Finally, we extend the analysis to the choice of sales outlets, exploring whether women farmers and men farmers differ in their reliance on farm-gate sales, local markets, or more distant markets. As outlets vary in competition, grading, and contract enforcement, they affect both price levels and volatility. Thus, who accesses which outlets has first-order implications for household income, risk, and incentives to invest. As commercialization in SSA remains in its early stages and markets are often thin with weak infrastructure, understanding where and how women sell farm products offers insight into the long-run process of price formation and market integration \citep{barrett_smallholder_2008}. 

The remainder of this paper is structured as follows: we present the country context in Section 2. In Section 3, we provide a detailed description of the data used in this study. Section 4 discusses the empirical method, and Section 5 presents the results. We provide alternative specifications in Section 6. Sections 7 and 8 conclude.

\section{Country context: Ethiopia, Nigeria, and Tanzania}

Country-specific context is essential for analyzing how gender influences agricultural market access across SSA. While many SSA countries are reliant on smallholder agriculture, institutional structures, cultural norms, and infrastructure vary, which can shape both the opportunities and constraints for market access and interactions. In this section, we discuss each country considered in our analysis and evaluate the similarities and differences across countries.  

First, we consider Ethiopia. Ethiopia’s agricultural sector is reliant on smallholder farming, with women farmers contributing significantly to crop production. According to \cite{palacios2017}, women account for 29\% of total labor in crop production. Despite their contributions, entrenched gender norms often restrict women's roles in market-related decision-making. In some regions, women may be prohibited from participating in public market negotiations, tangibly affecting market access. These constraints are amplified by limited land rights and low access to extension services. For example, while efforts have been made to promote joint land ownership to improve women’s bargaining power, decision-making over land use remains dominated by men in many households \citep{aguilar2015decomposition}. Thus, women farmers in Ethiopia face structural and institutional constraints that diminish their participation in agricultural commercialization.

Next, we turn to Nigeria. Nigeria offers a complex picture due to its diverse ethnic, cultural, and religious landscape. In some states, particularly in the southern part of the country, women play a large role in both agricultural production and commercialization. Women own and manage small farms, transport goods to markets, and operate as traders. Informal market networks provide a critical support structure, enabling women to access information, negotiate prices, and build social capital \citep{ajani2009gender}. However, in other states, particularly in the northern part of the country, socio-religious norms often influence women’s mobility and participation in public spaces. These restrictions lead to gendered divisions of labor where women may contribute to production but are excluded from direct sales, leading to lower returns and market participation \citep{ajani2009gender, worldbank2023gender}. Further, access to formal credit and land remains highly gendered, limiting entrepreneurial capacity among northern women farmers \citep{ajani2009gender, meagher_identity_2010}.

Finally, we consider Tanzania. Tanzania presents a gender-inclusive model, though challenges remain. Women contribute approximately 52\% of total crop labor \citep{palacios2017}, and they are active in both production and commercialization. Women often participate in agricultural cooperatives and market activities \citep{ellis_gender_2007}. In some regions, women sell at weekly markets and even engage in cross-border trade with neighboring countries. While cultural barriers still exist, particularly in more conservative rural communities, national gender policies and donor-supported initiatives have promoted women's empowerment in agriculture \citep{msonganzila2013gender}. Programs that offer training, market information, and group marketing have allowed some women to bypass middlemen and access better prices and more independence \citep{achandi_womens_2019}. Still, constraints like limited transport, poor storage, and low access to mobile technologies continue to disproportionately affect women, especially those in isolated areas \citep{meinzen-dick_womens_2019}.

Across all three countries, poor rural infrastructure remains a major constraint to market access, with disproportionate effects on women. In Ethiopia, remote regions suffer from inadequate roads and transportation services, limiting women's ability to reach urban markets. Women’s mobility is further constrained by household responsibilities and social expectations that discourage long-distance travel. As a result, they often sell to local middlemen at reduced prices, weakening their commercial returns \citep{minot2022commercialization}. Nigeria’s infrastructure landscape is more developed, and so roads, mobile money, and market facilities are generally accessible. In Tanzania, efforts to improve feeder roads and expand agricultural extension services have increased women’s engagement in cooperative marketing. However, the benefits tend to be concentrated in more developed regions, while women in remote districts still face significant hurdles \citep{msonganzila2013gender}.

Despite their differences, Ethiopia, Nigeria, and Tanzania share key characteristics that shape market access for women farmers: reliance on rainfed agriculture with high labor contributions from women, but persistent gender gaps in land rights, finance, and decision-making. Additionally, cultural and institutional norms often limit women’s economic roles. Together, these may influence commercialization and women's interactions with markets in country-specific ways. 

\section{Data and descriptive statistics}\label{sectdata}

We use nationally representative longitudinal household survey data from the Living Standard Measurement Study - Integrated Survey in Agriculture (LSMS-ISA) for Ethiopia, Nigeria, and Tanzania. These data are collected by the National Bureau of Statistics of each country with technical and logistical support from the World Bank. In each country, the surveys are implemented every other year, starting in Tanzania in 2008/09, in Nigeria in 2010/11, and in Ethiopia in 2011/12. In Ethiopia and Tanzania, the fourth and fifth waves of surveys are a refreshed panel, and they do not survey the same households and plots from the first three waves. Therefore, we use only the first three rounds of data to ensure comparability, covering the period from 2008 to 2016. The dataset includes detailed data on household and community characteristics, crop production, consumption, and marketing information. All the LSMS-ISA surveys follow a two-stage cluster sampling design, with households as sampling units and enumeration areas as clusters.\footnote{Further details on survey methodology and sampling procedures are described in the official guidelines and resources provided by the World Bank’s Living Standards Measurement Study (LSMS) \citep{world_bank_lsms_isa}.} 

As we estimate the relationship between farmers' gender and crop commercialization, our analysis is restricted to the sample of agricultural households. This is defined as the group of households that cultivated at least one crop in the 12 months preceding the survey.\footnote{As shown in Appendix~\ref{tableA2}\label{ref_to_tableA2}, around 90\% of households are rural households in all three countries.} Considering only these agricultural households gives us a panel of 2,118 agricultural households in Ethiopia, 2,189 agricultural households in Nigeria, and 1,704 agricultural households in Tanzania. Table~\ref{table1}\label{ref_to_table1} summarizes LSMS‑ISA panel sample sizes and survey years for Ethiopia, Nigeria, and Tanzania across three waves. 

\subsection{Key variables}

We have two key sets of variables in our analysis. First, we consider three measures of agricultural commercialization, and next, we consider several metrics of gender. 

\subsubsection{Agricultural commercialization}

Agricultural commercialization is our outcome variable of interest. We measure it with three key indicators: (1) participation in crop sales,\footnote{A household participation dummy equal to 1 if the household sold at least one crop in the past 12 months.} (2) the crop commercialization index (CCI), and (3) the cash-crop sales ratio. 

A binary indicator of crop sales can mask the underlying heterogeneity in sales participation. While it identifies whether a household sold any crop output, it does not capture details such as the quantity sold, the types of crops sold, or the location of sales. To more fully investigate these dynamics, we complement the binary indicator with two additional measures. The first of these is the crop commercialization index (CCI), which measures the share of harvest value sold. We follow \cite{carletto_agricultural_2017} to define the crop commercialization index (CCI) as:

\begin{equation}
\mathrm{CCI}_{c,i,t} = 
\frac{\text{Gross value of crop sales}_{c,i,t}}
{\text{Gross value of all crop harvested}_{c,i,t}}, 
\label{eq:cci}
\end{equation}

\noindent where $c$ indexes country, $i$ indexes household, and $t$ indexes survey wave. CCI values range between zero and one. A value of zero indicates pure subsistence, whereas a value of one indicates complete commercialization. 

In addition, we calculate the cash crop ratio, as neither the binary indicator nor the CCI indicates what sort of crops are sold. The cash crop ratio is the ratio of the value of cash crops sold to the value of total crop sales. This measure also ranges between zero and one, where a value of zero indicates that no cash crops are sold, and a value of one indicates that only cash crops are sold.  

Table~\ref{table2}\label{ref_to_table2} Panel A presents the summary statistics of all three crop commercialization variables. In all three countries, about half of the agricultural households sell some part of their crop production. In Ethiopia, 47\% of agricultural households sold crops, while in Nigeria, the share is 73\% of agricultural households. Crop commercialization also varies across countries, as the average CCI is 10\% in
Ethiopia, 28\% in Nigeria, and 26\% in Tanzania (Table~\ref{table2}\label{ref_to_table2} Panel A). Likewise, the average cash‑crop ratio is about 23\% in Ethiopia, 51\% in Nigeria, and 49\% in Tanzania.  

\subsubsection{Gender}

Our variable of interest is farmers' gender.  First, we use the gender of the household head,\footnote{The gender of the household head could change over time, which occurred in fewer than 100 households per country in our dataset.} which serves as a proxy for overall household leadership and decision-making, including control over resources and the commercialization process. Second, we calculate the share of land area managed by women members of the household, which reflects the extent of women’s direct involvement in agricultural production and plot-level management within the household. These two measures allow us to examine both household-level leadership and the degree of active participation by women as agricultural producers. Ideally, we would directly measure farmers’ gender at the plot level to provide more accurate insights into women’s participation in agricultural production and commercialization. Although our data include information about the gender of plot managers for each plot in every country, data on crop sales are only reported at the household level, not the plot level. As a result, we are unable to link sales from a specific plot to the gender of its manager, as we only have access to aggregated household sales.

Table~\ref{table2}\label{ref_to_table2} Panel B presents the summary statistics of gender variables. These are population-weighted averages across the three survey waves. Across three waves, about 20\% of agricultural households are headed by women, with country-specific figures of 18\% in Ethiopia, 11\% in Nigeria, and 23\% in Tanzania. The average share of women-managed land area varies across countries, with women farmers managing about 17\% of land in Ethiopia, 15\% in Nigeria, and 25\% in Tanzania.   

\subsection{Crop sales channels and locations} \label{sectypo}

The channels of crop sales in the LSMS‐ISA survey are elicited by asking about the primary point of crop harvest sales. We categorize the channels according to three dimensions: buyer type, market type, and sales location.  

First, we categorize crop sales based on buyer types. As the buyer-type labels in the data vary across countries and survey waves, we recode transaction-level buyer descriptors into harmonized categories to preserve comparability without collapsing heterogeneous partners into a single bin. To identify the gender differences in the choice of sales outlets, we construct three sub-categories of buyer types from transaction-level buyer descriptions. First, we group buyers into three mutually exclusive categories: (1) markets,\footnote{This includes any market-related sales, such as local markets, main markets, formal markets, or informal markets.} (2) individuals,\footnote{This includes sales to individual traders, neighbors, relatives, or other individual buyers.} and (3) non-market entities.\footnote{This includes sales to non-market entities that are not individuals, such as cooperatives, institutions, or other organizations.} Table~\ref{table3}\label{ref_to_table3} presents summary statistics for this category. 

Second, we categorize crop sales according to the market types, distinguishing between formal outlets and informal outlets.\footnote{This classification is based on the outlet names provided in the survey. For example, sales coded as formal include main markets, local markets, and other markets. Sales coded as informal include traders, roadside shops, neighbors, and relatives.} When sales outlets are missing or have ambiguous descriptors, we code them as ``unknown''. Our categorization may not fully capture the complexity of household marketing strategies, as a single household often sells different crops through a variety of outlets. Table~\ref{tableA1} Panel A in the Appendix presents the summary statistics for this category.

Third, we categorize crop sales into three categories using the geographic location of crop sales, which is recorded in the survey as where the transaction took place relative to the household's location: (1) within the village, (2) within the district, and (3) outside the district.\footnote{The survey does not provide exact definitions for these categories across the three countries. The ``within the village'' could refer to a range of distances. In Tanzania, for example, the guidance specifies that the location should reflect where the transaction occurred: if a buyer comes to the village to collect a crop, the sale is recorded as ``within the village'', even if the buyer’s office is elsewhere.} Table~\ref{tableA1} Panel B in the Appendix presents the summary statistics for this category. Across all three countries, most sales occur within the village, while outside-district sales are rare, with less than 15\% of farming households selling crops outside the district. 

We categorize crop sales in this way for several reasons. First, our categories of crop sales reflect potential economic mechanisms. The types of buyers capture bargaining power, pricing, and potentially even contracting features of markets, while the location of sales captures search and transport costs that shape market participation. Second, these categories map directly to plausible policy levers: road and market infrastructure affect the location margin (whether farmers can physically access markets) while trader regulation, licensing, and competition policy affect the counterparty margin (whether and which traders farmers can sell to, and under what conditions). Third, multiple categories provide internal validation and confirmation for our analysis: countries with higher ``formal market'' shares should also exhibit higher ``sold in the market'' shares. We observe this in our summary statistics, and it reduces the risk that any single coding choice drives findings. Fourth, the approach improves external validity and replicability: harmonized buckets can be applied consistently across LSMS-ISA waves and to other multi-country datasets despite idiosyncratic label differences. These constructs balance the need for cross-country harmonization with sufficient nuance to detect meaningful differences in market participation, while acknowledging that households often sell multiple crops through diverse outlets over time.

\subsection{Control covariates} 

We also include a set of control covariates in our estimation. Agricultural characteristics such as land area, access to irrigation, fertilizer use, transportation costs for crop sales, and exposure to weather shocks are considered, alongside household attributes like household size, dependency ratio, rural residence status, per-capita consumption expenditures, credit access, the share of consumption derived from own production, and distance to the nearest market. Additionally, we control for household head characteristics, including age, marital status, and educational attainment. These covariates are included in all empirical models. Table~\ref {tableA2}\label{ref_to_tableA2} in the Appendix presents the summary statistics for all control covariates for each wave and country.

\section{Empirical estimation strategy}

We estimate the relationship between gender and crop commercialization using panel data estimators in three countries. Specifically, we employ the two-way fixed effects (TWFE) estimator to control for unobserved household and time effects, and the two-way Mundlak (TWM) estimator, which permits coefficient estimation for time-invariant variables such as the gender of the household head by including household- and period-specific means of time-varying covariates. These approaches allow us to assess both "within" and "between" household associations across multiple waves of data. Standard errors are clustered at the household level for all regression models presented in this analysis.

We begin by using the two-way fixed effect estimator (TWFE), in which we regress each commercialization indicator on gender variables separately, controlling for time- and household-fixed effects. 

Let $Y_{it}$ be an indicator of crop commercialization of household $i$ at time $t$, $Gender_{it}$ be the gender of the head of household or the share of the women-managed land area in household $i$ at time $t$, and $X_{it}$ be the vector of variables that could affect crop sales by household $i$ at time $t$. Then, we can use Equation~\ref{eq1} to estimate the effects of gender variables on crop commercialization using the TWFE estimator. 

\begin{equation} \label{eq1}
Y_{it}= \alpha_1 Gender_{i} + \alpha_2 X_{it} + \mu_i + Time_t + \varepsilon_{it}.
\end{equation}

In Equation~\ref{eq1}, $\mu_i$ is the household fixed effects, $Time_t$ is the survey-year fixed effects, $\varepsilon_{it}$ is the idiosyncratic error term, and $\alpha_1$ is the coefficient of interest.

However, equation~\ref{eq1} is unable to estimate the effects of gender on commercialization as an individual's gender does not change over time.\footnote{While the household head can change over the three survey waves due to factors such as death or out-migration, this was an infrequent event in our sample, affecting fewer than 100 households per country.} Using the TWFE estimator would purge the effects of this gender dynamic. So, to address this, we use the two-way Mundlak (TWM) estimator. The TWM estimator was initially proposed by \cite{wooldridge_two-way_2021}, who has demonstrated its equivalence with TWFE. The equivalence between TWM and TWFE is achieved by including panel unit constant averages instead of panel-fixed effects and time-constant averages in place of time-fixed effects of all right-hand side variables in the estimating equations. Equation~\ref{eq2} provides the estimating equation for the TWM estimator. Instead of time- and household-fixed effects, we now include the vector of household-constant means of control covariates, $X_{i.}$, and the vector of time-constant means of control covariates, $X_{.t}$.

\begin{equation} \label{eq2}
Y_{it}= \beta_1 Gender_{i} + \beta_2 X_{it} + \beta_3 X_{i.} + \beta_4 X_{.t} + u_{it},
\end{equation}

\noindent where $Y_{it}$ denotes three commercialization indicators (whether a household sold crop produce, CCI, cash crop ratio) by household i at time t, $Gender_{i}$ denotes the household head’s gender or share of land managed by women members in household i, and $X_{it}$ denotes the vector of control variables that might affect the outcome of the sales for household i at time t, such as household size, dependent ratio, land size, etc. $X_{.i}$ is the vector of panel unit constant means of control covariates. $X_{.t}$ is the vector of time-constant means of control covariates.

To test the effect of the farmers’ gender on the selling positions in the market chain, we use the following specification:

\begin{equation} \label{eq3}
M_{it}= \beta_1 Gender_{i} + \beta_2 X_{it} + \beta_3 X_{i.} + \beta_4 X_{.t} + u_{it}, 
\end{equation}

\noindent where $M_{it}$ denotes three market categories by household i at time t, $Gender_{i}$ denotes the household head’s gender or share of land area managed by women members in the household $i$, and $X_{it}$ denotes the vector of control variables that might affect the outcome of the sales for household $i$ at time $t$. $X_{.i}$ is the vector of panel unit constant means of control covariates, and $X_{.t}$ is the vector of time-constant means of control covariates.

\section{Results} 
In this section, we present the main findings from our analysis across Ethiopia, Nigeria, and Tanzania. We use multiple indicators and econometric approaches to assess how gender dynamics within agricultural households relate to commercialization. The results reveal patterns of crop sales behavior, differences in market participation, and the impact of various household and farm characteristics, with a particular focus on gender-related differences observed across these three countries.
\subsection{Gender and commercialization}

Table~\ref{table4}\label{ref_to_table4} presents the relationship between farmers' gender, measured by an indicator for women-headed households and the share of land area managed by women, and three commercialization outcomes: crop sales participation in column 1, crop commercialization index in column 2, and cash crop sales ratio in column 3.

The results show that women farmers engage less in agricultural commercialization across all countries and measures examined. In Ethiopia, women-headed households are 6.6 percentage points less likely to participate in crop sales, have a crop commercialization index that is 3 percentage points lower, and a cash crop ratio that is 7.7 percentage points lower compared to men-headed households. Similarly, a doubling in the share of land managed by women corresponds with an 8 percentage point decrease in crop sales participation, a 3 percentage point decrease in the crop commercialization index, and a 6.8 percentage point drop in the cash crop ratio. All of these relationships are statistically significant at the 5\% significance level or below.  

Turning to Nigeria, a similar pattern emerges. Women-headed households are 8.9 percentage points less likely to participate in crop sales, have 5.9 percentage points lower crop commercialization index (CCI), and 7.3 percentage points lower cash crop ratio compared to men-headed households. All of these differences are statistically significant at the 5\% level. Moreover, the share of women-managed land area is also negatively associated with commercialization outcomes. Households with a higher proportion of women-managed land area are less likely to participate in crop sales, have lower CCI, and a lower cash crop ratio, with all relationships being statistically significant at the 1\% level. 

Last, in Tanzania, agricultural commercialization is negatively associated with women farmers, though the relationship is less robust than in Ethiopia and Nigeria. While we find that women-headed households are 4.6 percentage points less likely to participate in crop sales and have 5 percentage points smaller cash crop ratio than men-headed households, the CCI relationship is no longer significant at the 10\% level. When evaluating the share of women-managed land area, the associations with commercialization outcomes remain negative, but are not statistically significant at the 10\% level except for the cash crop ratio.

\subsection{Dissecting the gender effects on sales outlets} 

Table~\ref{table5}\label{ref_to_table5} explores the relationship between gender and the types of sales outlets. As discussed in section \ref{sectypo}, we disaggregate crop sales into three categories based on types: sold to markets, sold to individuals, and sold to non-market entities. Contrary to the common narrative, we do not find evidence that women farmers are less likely to sell to \textit{markets} and more likely to sell to friends or family. In fact, Table~\ref{table5}\label{ref_to_table5} shows that, conditional on selling crop products, for women-headed households, the results show differences in the types of buyers they sell to, particularly in Ethiopia and Nigeria. In Ethiopia, women-headed households are 7.1 percentage points more likely to sell to market buyers and 9.6 percentage points less likely to sell to individual buyers compared to men-headed households. There is no significant difference in sales to entities. In Nigeria, women-headed households are 11.9 percentage points more likely to sell to market buyers, while their likelihood of selling to individuals or entities does not significantly differ from that of men-headed households. By contrast, in Tanzania, there are no statistically significant differences in buyer types between women- and men-headed households. A higher share of women-managed land area is associated with significantly fewer sales to individual buyers, with no statistically significant association with sales to markets or non-market entities. Doubling the share of women-managed land area is associated with a higher likelihood of selling to markets but a lower likelihood of selling to individual buyers. Similarly, a greater share of women-managed land area is not statistically significantly associated with crop sales to non-market entities, except in Nigeria, where doubling the share of women-managed land area is associated with a 2.4 percentage point lower likelihood of selling to non-market entities. 

We also categorize crop sales based on market types (informal or formal markets) and the location of sales (within a village, within a district, or outside of the district). Appendix Table~\ref{tableA3}\label{ref_to_tableA3} presents the associations between farmers' gender and crop sales by market types. We find that in Nigeria, women-headed households are 10 percentage points more likely to sell to informal markets than men-headed households, and, as expected, the share of women-managed land area is also positively associated with crop sales to informal markets and negatively associated with crop sales to other outlets. Appendix Table~\ref{tableA4}\label{ref_to_tableA4} presents the association between farmers' gender and the location of crop sales. We do not find significant gender gaps in where transactions occur in Ethiopia and Tanzania. However, in Nigeria, a higher share of women-managed land area is associated with more sales within the village and fewer sales outside the district.

\subsection{Gender, commercialization, and home consumption}

Our results so far indicate that while women farmers in Sub-Saharan Africa are significantly less commercialized than men farmers, women are not necessarily less likely to sell to market outlets than men farmers. This raises the question: why are women farmers more disadvantaged in agricultural commercialization than men farmers, while they are not worse off than men in selling crops to markets? To address this puzzle, we investigate whether the relationship between gender and commercialization is mediated by self-consumption, the share of production consumed within the household. 
 
Table~\ref{table6}\label{ref_to_table6} presents the relationships between gender and commercialization, accounting for self-consumption. Results show that self-consumption is significantly negatively associated with the probability of selling crops, and this relationship is stronger among women-headed households than men-headed households. This suggests that households prioritizing their own consumption are less likely to sell crops. To investigate this further, we test two different measures of gender involvement: a binary indicator for women-headed households (presented in Panel A) and a continuous variable for the share of land managed by women (presented in Panel B). The models are estimated separately for Ethiopia, Nigeria, and Tanzania.

In Ethiopia (presented in Panel A), we observe that higher self-consumption is negatively correlated with crop sales. Households with greater self-consumption are 15.2 percentage points less likely to participate in crop sales. The interaction term suggests that women-headed households with high self-consumption are disproportionately disadvantaged in market sales, with an additional reduction of 15.1 percentage points relative to men-headed households. Results are similar in Nigeria, where self-consumption lowers sales by 36.2 percentage points, and the interaction term shows that women-headed households experience an additional 15.2 percentage-point reduction in commercialization. In both contexts, women-headed households appear doubly constrained: they face gender-specific disadvantages in commercialization, and these disadvantages are amplified when a larger share of production is allocated to self-consumption. In contrast, in Tanzania, neither self-consumption nor its interaction with women's headship is statistically significant, consistent with the less robust gender effects found in Table~\ref{table4}\label{ref_to_table4}. 

Considering a different dynamic, the share of women-managed land (presented in Panel B), we find that again, Ethiopia and Nigeria show strong negative moderating effects of self-consumption. In Tanzania, however, no significant effects are observed.

\section{Alternative specifications} 

Next, we consider alternative specifications of these baseline results. We employ two complementary specifications to investigate the relationship between farmers’ gender and commercialization. First, we proxy farmers’ gender with the gender of the agricultural decision-maker, which is identified by authority over harvest allocation. We do this, as it most directly links production to sales. Next, we pool data from the three countries to estimate average regional trends.\footnote{We also estimate pooled OLS models as an alternative to our baseline approach to assess robustness. The results of pooled OLS models are displayed in the Appendix Table~\ref{tableA5} but are not in the main text.} 

\subsection{Decision-makers' gender and commercialization}

To further investigate the gender dynamics of commercialization, we proxy agricultural decision-making with an indicator that captures the gender of the individual who holds authority over the allocation of the harvest. This decision is most proximate to commercialization and links production to income and, ultimately, to food security. This measure aligns with established empowerment proxies. We thus treat it as a rough proxy for the gender of the agricultural decision-maker.\footnote{We acknowledge, however, that it does not fully capture asset ownership, labor allocation, or broader expenditure control.}

Households with a woman head or more women members involved in agricultural management exhibit lower commercialization. However, when we proxy the gender of the agricultural decision-maker using authority over harvest allocation and condition on observables, we find a negative relationship as observed in our primary results, but we do not observe a statistically significant association with sales, CCI, or the cash crop ratio in Table~\ref{table7}\label{ref_to_table7}. This pattern suggests that gender gaps in commercialization are driven more by factors at the production and consumption level—such as resource endowments and decisions about allocating harvest for home use, rather than by decision-making dynamics during the market transaction itself. 

\subsection{Pooled country analysis}

Finally, to test whether an overall association exists across the region and to establish a benchmark for comparison, we pool the data from all three countries. This pooled estimate serves to quantify the average relationship, against which the country-specific results can be more clearly interpreted. Considering the similarity of survey structure and data across countries, the pooled analysis complements the country-by-country estimates and is suggestive of general regional behavior. Results in Table~\ref{table8}\label{ref_to_table8} show consistent negative associations between farmers' gender and commercialization, after controlling for household demographics, farm characteristics, market access, climate shocks, and country fixed effects. Women-headed households are 6.6 percentage points less likely to sell, with a 2.9 point lower CCI and a 6.2 point lower cash crop share compared with the men-headed households. A higher share of women-managed land is 8.6 percentage points less likely to sell, with a 3.3 point lower CCI and an 8.4 point lower cash crop share.

\section{Discussion} 

Taken together, our results document sizable and systematic gender gaps in agricultural commercialization in Ethiopia and Nigeria, with less consistent patterns in Tanzania. Households in which women hold key roles, either as heads or as managers of a larger share of land, are consistently less likely to participate in crop sales, exhibit lower CCI, and devote smaller shares of output to cash crops. Although our empirical models include a rich set of controls, these estimates should be interpreted as associations rather than causal effects. Even with these limitations, the pattern is consistent with women facing tighter constraints in accessing the inputs, credit, labor, information, and marketing networks required to shift production toward market-oriented and cash-crop activities.

Decomposition by buyer type, market type, and sales location provides additional evidence. Contrary to the common narrative that women prefer or are restricted to interpersonal transactions, we find that, conditional on selling, women-headed households in Ethiopia and Nigeria are actually more likely to sell to market buyers. In Tanzania, there is no statistically significant difference in market engagement by gender. In Nigeria specifically, women’s sales are more likely to occur through informal markets and within the village, rather than outside the district, and households with a greater share of women-managed land area report fewer sales to individual buyers. These results suggest that women's limited engagement with formal or distant markets is driven less by reluctance and more by constraints related to mobility, safety, time, or liquidity. Instead of avoiding markets, women tend to connect through local and less intermediated channels, reflecting these practical barriers rather than a preference for non-market transactions.

The analysis of self-consumption offers a plausible mechanism for the observed gender gaps in commercialization. In Ethiopia and Nigeria, self-consumption is strongly and negatively associated with market participation, and the gap is significantly larger for women-headed households. This pattern is also evident when gender is measured by the share of land area managed by women, which shows similar moderating effects on commercialization outcomes. These results imply that when a higher share of output is allocated to home use (potentially reflecting women’s primary role in household consumption decisions), the residual marketable surplus falls, ultimately depressing both participation and specialization in cash crops. This mechanism helps reconcile the observation that women sell less overall, even though, when they do sell, they do not systematically avoid market outlets. In Tanzania, the absence of a significant relationship between self-consumption and commercialization aligns with the generally less robust gender patterns observed in the baseline results.

Alternative specifications support these interpretations. When gender is proxied by authority over harvest allocation, which is the decision most closely tied to commercialization, the gender of the agricultural decision-maker is negatively associated with commercialization, although this relationship is not statistically significant. This attenuation suggests that endowments, production choices, liquidity, and consumption needs upstream of the sale decision likely account for much of the observed gap, rather than bargaining, relationships, and interactions at the point of sale. The pooled analysis indicates that, on average, the challenges faced by women farmers in the process of commercialization are a significant regional phenomenon. While the paper’s main finding is that the magnitudes of these gaps are highly context-specific, the pooled results suggest that the underlying constraints are widespread. This reinforces the need for policy attention on this issue across Sub-Saharan Africa, even as specific interventions must be tailored to local realities.

Taken together, this evidence points to policies that relax practical constraints faced by women rather than attempting to redirect sales channels. First, interventions that address the trade-off between self-consumption and commercialization, such as improved on-farm storage, reliable access to staple foods, seasonal safety nets, and short-term credit, may be consequential for women-managed land area and farms. Second, reducing proximity and mobility frictions through safer transport, village-level aggregation points, and trustworthy intermediaries can facilitate participation beyond local outlets. 

\subsection{Heterogeneity and limitations}

It is worth noting that cross-country variation, particularly Tanzania’s muted patterns, underscores that gender gaps in commercialization are not uniform and likely depend on crop mix, market structure, and local norms. The analysis remains observational and relies on proxies, such as women’s headship, women-managed land share, and authority over harvest allocation, which do not fully capture control over assets, labor, or income. Self-consumption may be endogenous to shocks and preferences, and the classification of ``informal'' outlets may obfuscate within-category heterogeneity. 

Ultimately, we conclude that women are not less ``market-oriented'' though they are less commercialized. We instead conclude that women farmers have more constrained access to markets than men. Reducing the frictions that disproportionately affect women offers a promising route to narrow gender gaps in commercialization while safeguarding food security and household welfare.

\section{Conclusion}

Using three waves of LSMS-ISA data from Ethiopia, Nigeria, and Tanzania, we document robust associations between women’s roles in the household and lower agricultural commercialization, particularly in Ethiopia and Nigeria. Women-headed households and those with greater shares of women-managed land sell less, have lower CCI, and allocate smaller sales shares to cash crops. In both Ethiopia and Nigeria, women who sell are more likely to use market buyers, though sales in Nigeria are further concentrated in proximate and informal markets. This suggests that there exists stronger mobility frictions. A key mechanism is the consumption–sales trade-off: self-consumption is strongly negatively associated with market participation in both Ethiopia and Nigeria, with larger penalties among women-headed households and where women manage more land. An alternative specification that proxies the gender of the agricultural decision-maker through harvest-allocation authority yields no statistically significant relationship with commercialization once observables are controlled, suggesting that upstream differences in endowments, production choices, liquidity, and consumption needs account for much of the observed gap. Pooled estimates that combine data from all three countries confirm the main results.

While our estimates are not causal, they are still informative for research and policy. First, these patterns provide important signals. They identify populations, such as women-headed or women-managed farms, and conditions, such as high self-consumption, where commercialization is consistently low. This can support targeting further measurement and program piloting. Second, they generate testable hypotheses about mechanisms for investigation in further research. Third, the cross-country consistency and robustness across multiple outcomes and measures of gender provide information for policy design: even if the magnitudes are biased by unobservables, the direction and relative ranking of constraints are likely informative. Finally, many plausible errors, such as classical measurement error in gender proxies or sales, would attenuate coefficients, implying that the true relationships could be at least as strong as those we document. Thus, associative evidence can productively guide where to invest scarce identification resources and inform policy moving forward. 

Based on our findings, policies that encourage commercialization and market engagement should target the constraints that bind women rather than attempting to re-route sales channels. Priorities include: (1) easing the self-consumption–commercialization trade-off through improved storage, seasonal safety nets, and short-term credit; and (2) lowering proximity and mobility frictions through safer transport, village-level aggregation points, and trustworthy intermediaries. Such measures expand women’s commercialization choices and opportunities while safeguarding household food security and ensuring household welfare.

The limitations remain with this work. Outcomes are measured at the household level, gender proxies are imperfect, and self-consumption may be endogenous to shocks and preferences. Future work should link plot- and transaction-level data to decision-makers and work to identify which constraints most effectively relax women’s bottlenecks. 

As commercial agriculture grows in SSA, embedding gender-responsive design in market infrastructure and institutions is essential to expand participation, raise productivity, and ensure that the gains from market growth are shared within households and across communities.

\clearpage
\newpage
\section*{\raggedright Tables}

\begin{table}[htbp]
\centering
\begin{threeparttable}
\caption{Sample sizes by country and wave \textit{(Return to page~\pageref{ref_to_table1})}}
    \label{table1}
    \begin{tabular}{@{\extracolsep{5pt}}l*{6}{c}}
      \toprule
      & \multicolumn{2}{c}{Wave 1} & \multicolumn{2}{c}{Wave 2} & \multicolumn{2}{c}{Wave 3} \\
    \cmidrule(l){2-3} \cmidrule(l){4-5} \cmidrule(l){6-7}
      
      Country & Survey  & Number of & Survey  & Number of & Survey & Number of \\
       & year  & households & year  & households & year & households \\      
      \midrule
      Ethiopia & 2011/12  & 2,476 & 2013/14  & 2,446 & 2015/16  & 2,317 \\
      Nigeria  & 2010/11  & 2,326 & 2012/13  & 2,340 & 2015/16  & 2,363 \\
      Tanzania & 2008/09  & 1,802 & 2010/11  & 1,853 & 2012/13  & 2,009 \\
      \bottomrule
   \end{tabular}
   \begin{tablenotes}[para]
  \footnotesize
  \item Notes: All survey waves are nationally representative, except for the 2011/12 wave of the Ethiopian panel, which is representative of rural and small-town areas only. All households in the sample are agricultural households that reported cultivating land in the last 12 months.
  \end{tablenotes}
  \end{threeparttable}
  \end{table}

\begin{sidewaystable}[htbp] 
\centering
\begin{threeparttable}
\def\sym#1{\ifmmode^{#1}\else\(^{#1}\)\fi}
\caption{Summary of key variables \textit{(Return to page~\pageref{ref_to_table2})}}
\label{table2}
\begin{tabular}{@{\extracolsep{5pt}}l*{9}{c}}
\toprule
                    &\multicolumn{3}{c}{Ethiopia}          &\multicolumn{3}{c}{Nigeria}           &\multicolumn{3}{c}{Tanzania}          \\\cmidrule(lr){2-4}\cmidrule(lr){5-7}\cmidrule(lr){8-10}
                    &\multicolumn{1}{c}{Wave 1}&\multicolumn{1}{c}{Wave 2}&\multicolumn{1}{c}{Wave 3}&\multicolumn{1}{c}{Wave 1}&\multicolumn{1}{c}{Wave 2}&\multicolumn{1}{c}{Wave 3}&\multicolumn{1}{c}{Wave 1}&\multicolumn{1}{c}{Wave 2}&\multicolumn{1}{c}{Wave 3}\\
\midrule
\emph{\textbf{A. Outcome variables}}&            &            &            &            &            &            &            &            &            \\
Sold any crops (1=yes)&        0.47&        0.53&        0.48&        0.60&        0.56&        0.73&        0.58&        0.60&        0.58\\
                    &      (0.50)&      (0.50)&      (0.50)&      (0.49)&      (0.50)&      (0.44)&      (0.49)&      (0.49)&      (0.49)\\
CCI                 &        0.20&        0.10&        0.07&        0.25&        0.25&        0.33&        0.24&        0.28&        0.25\\
                    &      (0.35)&      (0.17)&      (0.15)&      (0.33)&      (0.32)&      (0.34)&      (0.32)&      (0.34)&      (0.33)\\
Cash crop ratio     &        0.21&        0.25&        0.22&        0.49&        0.44&        0.60&        0.58&        0.60&        0.55\\
                    &      (0.38)&      (0.40)&      (0.39)&      (0.47)&      (0.46)&      (0.45)&      (0.49)&      (0.49)&      (0.50)\\
\vspace{0.1em} \\ \emph{\textbf{B. Gender variables}}&            &            &            &            &            &            &            &            &            \\
Woman household head (1=yes)&        0.17&        0.18&        0.19&        0.10&        0.11&        0.13&        0.22&        0.22&        0.24\\
                    &      (0.37)&      (0.38)&      (0.40)&      (0.30)&      (0.31)&      (0.34)&      (0.42)&      (0.42)&      (0.43)\\
Share of women-managed  land area&        0.17&        0.17&        0.18&        0.15&        0.15&        0.16&        0.24&        0.24&        0.26\\
                    &      (0.37)&      (0.38)&      (0.38)&      (0.35)&      (0.36)&      (0.36)&      (0.42)&      (0.43)&      (0.43)\\
\midrule
Observations        &       2,118&       2,118&       2,118&       2,189&       2,189&       2,189&       1,704&       1,704&       1,704\\
\bottomrule
\end{tabular}
\begin{tablenotes}[para]
\footnotesize
\item Notes: Point estimates are sample means. Standard errors are in parentheses. The crop commercialization index (CCI) is the ratio of the gross value of crops sold to the gross value of crops harvested per season or year. CCI ranges between 0 (subsistence) and 1 (complete commercialization) (see \cite{carletto_agricultural_2017}). The cash crop ratio is calculated as the proportion of crops sold that come from cash crops.
\end{tablenotes}
\end{threeparttable}
  \end{sidewaystable}

\begin{sidewaystable}[htbp]
\centering
\begin{threeparttable}

\def\sym#1{\ifmmode^{#1}\else\(^{#1}\)\fi}
\caption{Crop sales channels: buyer type  (sellers only)\textit{(Return to page~\pageref{ref_to_table3})}}
\label{table3}
\begin{tabular}{@{\extracolsep{5pt}}l*{9}{c}}
\toprule
                    &\multicolumn{3}{c}{Ethiopia}          &\multicolumn{3}{c}{Nigeria}           &\multicolumn{3}{c}{Tanzania}          \\\cmidrule(lr){2-4}\cmidrule(lr){5-7}\cmidrule(lr){8-10}
                    &\multicolumn{1}{c}{Wave 1}&\multicolumn{1}{c}{Wave 2}&\multicolumn{1}{c}{Wave 3}&\multicolumn{1}{c}{Wave 1}&\multicolumn{1}{c}{Wave 2}&\multicolumn{1}{c}{Wave 3}&\multicolumn{1}{c}{Wave 1}&\multicolumn{1}{c}{Wave 2}&\multicolumn{1}{c}{Wave 3}\\
\midrule
\emph{\textbf{A. Buyer type}}   &            &            &            &            &            &            &            &            &            \\
Sold at any market (1=yes)&        0.54&        0.68&        0.65&        0.72&        0.74&        0.66&        0.22&        0.64&        0.66\\
                    &      (0.50)&      (0.47)&      (0.48)&      (0.45)&      (0.44)&      (0.47)&      (0.41)&      (0.48)&      (0.47)\\
Sold to individual (1=yes)&        0.26&        0.29&        0.31&        0.28&        0.30&        0.25&        0.53&        0.17&        0.14\\
                    &      (0.44)&      (0.45)&      (0.46)&      (0.45)&      (0.46)&      (0.43)&      (0.50)&      (0.37)&      (0.35)\\
Sold to entity (1=yes)&        0.05&        0.03&        0.04&        0.05&        0.06&        0.04&        0.33&        0.16&        0.13\\
                    &      (0.23)&      (0.16)&      (0.20)&      (0.22)&      (0.24)&      (0.19)&      (0.47)&      (0.37)&      (0.34)\\
Sold details unknown (1=yes)&        0.18&        0.04&        0.04&        0.03&        0.02&        0.11&        0.02&        0.00&        0.01\\
                    &      (0.39)&      (0.19)&      (0.19)&      (0.17)&      (0.13)&      (0.31)&      (0.15)&      (0.04)&      (0.10)\\
\midrule
Observations        &         996&       1,123&       1,007&       1,307&       1,219&       1,603&         986&       1,023&         984\\
\bottomrule
\end{tabular}
\begin{tablenotes}[para]
\footnotesize
\item Notes: Point estimates are sample means. Standard errors are in parentheses. Other category groups can be found in the Appendix Table~\ref{tableA1}\label{ref_to_tableA1}. 
\end{tablenotes}
\end{threeparttable}
\end{sidewaystable}

\begin{table}[htbp]
\centering
\begin{threeparttable}
\caption{Effects of gender on commercialization: TWM results \textit{(Return to page~\pageref{ref_to_table4})}}
\label{table4}
\begin{tabular}{@{\extracolsep{5pt}}l*{3}{c}}
\toprule
                    &\multicolumn{1}{c}{(1)}&\multicolumn{1}{c}{(2)}&\multicolumn{1}{c}{(3)}\\
                    &\multicolumn{1}{c}{Sales (HH)}&\multicolumn{1}{c}{CCI}&\multicolumn{1}{c}{Cash Crop Ratio}\\
\midrule \multicolumn{4}{l}{\textbf{Panel A: Ethiopia}} \\
\addlinespace
Woman household head (1=yes)&      -0.066\sym{**} &      -0.030\sym{**} &      -0.077\sym{***}\\
                    &     (0.030)         &     (0.013)         &     (0.023)         \\
\addlinespace
Observations        &        6072         &        6072         &        6072 \\

\addlinespace
Share of women-managed  land area&      -0.080\sym{***}&      -0.030\sym{***}&      -0.068\sym{***}\\
                     &     (0.027)         &     (0.012)         &     (0.021)         \\
\addlinespace

Observations        &        6,065         &        6,065         &        6,065         \\
\midrule \multicolumn{4}{l}{\textbf{Panel B: Nigeria}} \\
\addlinespace
Woman household head (1=yes)&      -0.089\sym{**} &      -0.059\sym{**} &      -0.073\sym{**} \\
                   &     (0.036)         &     (0.027)         &     (0.036)         \\
\addlinespace

Observations        &        6394         &        6394         &        6394         \\

\addlinespace
Share of women-managed  land area&      -0.099\sym{***}&      -0.052\sym{***}&      -0.112\sym{***}\\
                     &     (0.025)         &     (0.018)         &     (0.025)         \\
\addlinespace

Observations        &        6,387         &        6,387         &        6,387         \\

\midrule \multicolumn{4}{l}{\textbf{Panel C: Tanzania}} \\
\addlinespace
Woman household head (1=yes)&      -0.046\sym{*}  &      -0.015         &      -0.050\sym{*}  \\
                   &     (0.030)         &     (0.019)         &     (0.030)         \\
\addlinespace

Observations        &        5003         &        5003         &        5003         \\

\addlinespace
Share of women-managed  land area&      -0.042         &      -0.014         &      -0.048\sym{*}  \\
                    &     (0.027)         &     (0.016)         &     (0.026)         \\
\addlinespace

Observations        &        4,750         &        4,750         &        4,750         \\
\bottomrule
\end{tabular}
\begin{tablenotes}[para]
\footnotesize
\item Notes: Standard errors clustered at the household level are in parentheses. All regressions include a standard set of control variables for household demographics (size, dependency ratio, head’s age, marital status, and education), agricultural and economic characteristics (land size, irrigation, loan access, and distance to market), and exposure to climatic shocks. CCI represents the crop commercialization index, calculated as the ratio of the gross value of crops sold to the gross value of crops produced per season or year. The cash crop ratio is calculated as the proportion of output sold that comes from cash crops. Level of statistical significance: \sym{*} \(p<0.10\), \sym{**} \(p<0.05\), \sym{***} \(p<0.01\).
\end{tablenotes}
\end{threeparttable}
\end{table}

\begin{table}[htbp]
\centering
\begin{threeparttable}
\def\sym#1{\ifmmode^{#1}\else\(^{#1}\)\fi}
\caption{Effects of gender on different buyer types \textit{(Return to page~\pageref{ref_to_table5})}}
\label{table5}
\begin{tabular}{@{\extracolsep{5pt}}l*{3}{c}}
\toprule
                    & \multicolumn{1}{c}{(1)} & \multicolumn{1}{c}{(2)} & \multicolumn{1}{c}{(3)} \\
                    & \multicolumn{1}{c}{Sold to Market} & \multicolumn{1}{c}{Sold to Individual} & \multicolumn{1}{c}{Sold to Entity} \\
\midrule
\multicolumn{4}{l}{\textbf{Panel A: Ethiopia}} \\
Woman household head (1=yes)&       0.071\sym{*}  &      -0.096\sym{**} &      -0.021         \\
                    &     (0.038)         &     (0.038)         &     (0.014)         \\
\addlinespace
Observations        &        2977         &        2977         &        2977         \\
\addlinespace
Share of women-managed  land area&       0.038         &      -0.059\sym{*}  &      -0.014         \\
                    &     (0.035)         &     (0.035)         &     (0.012)         \\
\addlinespace
Observations        &        2973         &        2973         &        2973         \\
\midrule
\multicolumn{4}{l}{\textbf{Panel B: Nigeria}} \\
Woman household head (1=yes)&       0.119\sym{***}&      -0.048         &      -0.020         \\
                    &     (0.041)         &     (0.038)         &     (0.022)         \\
\addlinespace
Observations        &        4045         &        4045         &        4045         \\
\addlinespace
Share of women-managed  land area&       0.102\sym{***}&      -0.063\sym{**} &      -0.024\sym{*}  \\
                    &     (0.029)         &     (0.028)         &     (0.014)         \\
\addlinespace
Observations        &        4039         &        4039         &        4039         \\
\midrule
\multicolumn{4}{l}{\textbf{Panel C: Tanzania}} \\
Woman household head (1=yes)&       0.003         &      -0.032         &       0.012         \\
                    &     (0.034)         &     (0.030)         &     (0.029)         \\
\addlinespace                    
Observations        &        2945         &        2945         &        2945         \\
\addlinespace
Share of women-managed  land area&      -0.023         &       0.008         &       0.004         \\
                    &     (0.031)         &     (0.030)         &     (0.027)         \\
\addlinespace
Observations        &        2833         &        2833         &        2833         \\
\bottomrule
\end{tabular}

\begin{tablenotes}[para]
\footnotesize
\item Notes: Standard errors clustered at the household level are in parentheses. ``Sold details unknown'' is the reference category. All regressions include a standard set of control variables for household demographics (size, dependency ratio, head’s age, marital status, and education), agricultural and economic characteristics (land size, irrigation, loan access, and distance to market), and exposure to climatic shocks. Level of statistical significance: \sym{*} \(p<0.10\), \sym{**} \(p<0.05\), \sym{***} \(p<0.01\). 
\end{tablenotes}
\end{threeparttable}
\end{table}

\begin{table}[htbp]
\centering
\begin{threeparttable}
\def\sym#1{\ifmmode^{#1}\else\(^{#1}\)\fi}
\caption{Effects of gender and self-consumption on commercialization \textit{(Return to page~\pageref{ref_to_table6})}}
\label{table6}
\begin{tabular}{@{\extracolsep{5pt}}l*{3}{c}}
\toprule
&\multicolumn{3}{c}{Dependent Variable: Sales(HH)}\\
\cmidrule(lr){2-4}
                    &\multicolumn{1}{c}{(1)}&\multicolumn{1}{c}{(2)}&\multicolumn{1}{c}{(3)}\\
                    &\multicolumn{1}{c}{Ethiopia}&\multicolumn{1}{c}{Nigeria}&\multicolumn{1}{c}{Tanzania}\\
\midrule
\multicolumn{4}{l}{\textbf{Panel A: Woman Head × Self-Consumption}} \\
\addlinespace
Woman household head (1=yes)&       0.050         &      -0.032         &      -0.031         \\
                   &     (0.054)         &     (0.037)         &     (0.032)         \\
\addlinespace
Self-Consumption    &      -0.152\sym{***}&      -0.362\sym{***}&      -0.022         \\
                     &     (0.027)         &     (0.025)         &     (0.020)         \\
\addlinespace
Woman household head (1=yes) $\times$ Self-Consumption&      -0.151\sym{**} &      -0.152\sym{***}&      -0.034         \\
                     &     (0.060)         &     (0.049)         &     (0.043)         \\
\addlinespace

Observations        &    6,072         &    6,394         &    5,003        \\
\midrule
\multicolumn{4}{l}{\textbf{Panel B: Women Plot Share × Self-Consumption}} \\
\addlinespace
Share of women-managed  land area&       0.061         &      -0.027         &      -0.043         \\
                    &     (0.053)         &     (0.027)         &     (0.028)         \\
\addlinespace
Self-Consumption    &      -0.147\sym{***}&      -0.344\sym{***}&       0.012         \\
                     &     (0.027)         &     (0.026)         &     (0.022)         \\
\addlinespace
Share of women-managed  land area $\times$ Self-Consumption&      -0.181\sym{***}&      -0.151\sym{***}&      -0.014         \\
                    &     (0.060)         &     (0.046)         &     (0.045)         \\
\addlinespace

Observations        &    6,065        &    6,387         &    4,750         \\
\bottomrule
\end{tabular}
\begin{tablenotes}[para]
\footnotesize
\item Notes: Standard errors clustered at the household level are in parentheses. Self-consumption refers to the share of production consumed by the household. All regressions include a standard set of control variables for household demographics (size, dependency ratio, head’s age, marital status, and education), agricultural and economic characteristics (land size, irrigation, loan access, and distance to market), and exposure to climatic shocks. Level of statistical significance: \sym{*} \(p<0.10\), \sym{**} \(p<0.05\), \sym{***} \(p<0.01\). 
\end{tablenotes}
\end{threeparttable}
\end{table}


\begin{table}[htbp]\centering
\begin{threeparttable}
\def\sym#1{\ifmmode^{#1}\else\(^{#1}\)\fi}
\caption{Decision-makers' gender on commercialization \textit{(Return to page~\pageref{ref_to_table7})}}
\label{table7}
\begin{tabular}{@{\extracolsep{5pt}}l*{3}{c}}
\toprule
                    &\multicolumn{1}{c}{(1)}&\multicolumn{1}{c}{(2)}&\multicolumn{1}{c}{(3)}\\
                    &\multicolumn{1}{c}{Sales (HH)}&\multicolumn{1}{c}{CCI}&\multicolumn{1}{c}{Cash Crop Ratio}\\
\midrule \multicolumn{4}{l}{\textbf{Panel A: Ethiopia}} \\ 
Share of women decision-makers      &       0.000         &      -0.024         &      -0.013         \\
                    &         (.)         &     (0.023)         &     (0.047)         \\
\addlinespace
Observations        &        1790         &        1790         &        1790         \\
\midrule \multicolumn{4}{l}{\textbf{Panel B: Nigeria}} \\ 
Share of women  decision-makers      &      -0.019         &      -0.020         &      -0.015         \\
                     &     (0.028)         &     (0.020)         &     (0.027)         \\
\addlinespace

Observations        &        3880         &        3880         &        3880         \\
\midrule \multicolumn{4}{l}{\textbf{Panel C: Tanzania}} \\ 
Share of women  decision-makers      &      -0.060         &      -0.042         &      -0.060         \\

 &     (0.042)         &     (0.026)         &     (0.042)         \\
 \addlinespace
Observations        &        1443         &        1443         &        1443         \\

\bottomrule 
\end{tabular}

\begin{tablenotes}[para]
\footnotesize
\item Notes: Standard errors clustered at the household level are in parentheses. All regressions include a standard set of control variables for household demographics (size, dependency ratio, head’s age, marital status, and education), agricultural and economic characteristics (land size, irrigation, loan access, and distance to market), and exposure to climatic shocks. The share of women decision-makers indicates the proportion of decision-makers within the household who are women, regarding the use of harvested crops. CCI represents the crop commercialization index, calculated as the ratio of the gross value of crops sold to the gross value of crops produced per season or year.  The cash crop ratio is calculated as the proportion of output sold that comes from cash crops. Level of statistical significance: \sym{*} \(p<0.10\), \sym{**} \(p<0.05\), \sym{***} \(p<0.01\). 
\end{tablenotes}
\end{threeparttable}
\end{table}


\begin{table}[htbp]
\centering
\begin{threeparttable}
\def\sym#1{\ifmmode^{#1}\else\(^{#1}\)\fi}
\caption{Country pooled effects of gender on commercialization \textit{(Return to page~\pageref{ref_to_table8})}}
\label{table8}
\begin{tabular}{@{\extracolsep{5pt}}l*{3}{c}}
\toprule
                    &\multicolumn{1}{c}{(1)}&\multicolumn{1}{c}{(2)}&\multicolumn{1}{c}{(3)}\\
                    &\multicolumn{1}{c}{Sales (HH)}&\multicolumn{1}{c}{CCI}&\multicolumn{1}{c}{Cash Crop Ratio}\\
\midrule
Woman household head (1=yes)&      -0.066\sym{***}&      -0.029\sym{**} &      -0.062\sym{***}\\
                    &     (0.018)         &     (0.011)         &     (0.016)         \\     
\addlinespace

Observations        &       17,469         &       17,469         &       17,469         \\
\midrule 
Share of women-managed  land area&      -0.086\sym{***}&      -0.033\sym{***}&      -0.084\sym{***}\\
                    &     (0.015)         &     (0.009)         &     (0.014)         \\
\addlinespace
                  
Observations        &       17,202         &       17,202         &       17,202         \\
\bottomrule
\end{tabular}

\begin{tablenotes}[para]
\footnotesize
\item Notes: Standard errors clustered at the household level are in parentheses. Country-specific fixed effects are included. All regressions include a standard set of control variables for household demographics (size, dependency ratio, head’s age, marital status, and education), agricultural and economic characteristics (land size, irrigation, loan access, and distance to market), and exposure to climatic shocks. CCI represents the crop commercialization index, calculated as the ratio of the gross value of crops sold to the gross value of crops produced per season or year.  The cash crop ratio is calculated as the proportion of output sold that comes from cash crops. Level of statistical significance: \sym{*} \(p<0.10\), \sym{**} \(p<0.05\), \sym{***} \(p<0.01\). 
\end{tablenotes}
\end{threeparttable}
\end{table}

\newpage
\clearpage
\begin{spacing}{1}
\bibliographystyle{apalike}
\bibliography{main.bib}
\end{spacing}


\newpage
\appendix
\section*{\raggedright Appendix}

\setcounter{table}{0} 
\renewcommand{\thetable}{A\arabic{table}}

\begin{sidewaystable}[htbp]\centering
\begin{threeparttable}
\def\sym#1{\ifmmode^{#1}\else\(^{#1}\)\fi}
\caption{Other crop sales channels (sellers only)}
\label{tableA1}
\begin{tabular}{@{\extracolsep{5pt}}l*{9}{c}}
\toprule
                    &\multicolumn{3}{c}{Ethiopia}          &\multicolumn{3}{c}{Nigeria}           &\multicolumn{3}{c}{Tanzania}          \\\cmidrule(lr){2-4}\cmidrule(lr){5-7}\cmidrule(lr){8-10}
                    &\multicolumn{1}{c}{Wave 1}&\multicolumn{1}{c}{Wave 2}&\multicolumn{1}{c}{Wave 3}&\multicolumn{1}{c}{Wave 1}&\multicolumn{1}{c}{Wave 2}&\multicolumn{1}{c}{Wave 3}&\multicolumn{1}{c}{Wave 1}&\multicolumn{1}{c}{Wave 2}&\multicolumn{1}{c}{Wave 3}\\
\midrule
\vspace{0.1em} \\ \emph{\textbf{A. Market type}}&            &            &            &            &            &            &            &            &            \\
Sold in formal market (1=yes)&        0.34&        0.42&        0.43&        0.16&        0.15&        0.16&        0.22&        0.63&        0.65\\
                    &      (0.47)&      (0.49)&      (0.50)&      (0.37)&      (0.36)&      (0.37)&      (0.41)&      (0.48)&      (0.48)\\
Sold in informal market (1=yes)&        0.47&        0.57&        0.54&        0.81&        0.85&        0.73&        0.77&        0.30&        0.25\\
                    &      (0.50)&      (0.50)&      (0.50)&      (0.39)&      (0.36)&      (0.44)&      (0.42)&      (0.46)&      (0.43)\\
Sold in other type of market (1=yes)&        0.04&        0.02&        0.02&        0.05&        0.06&        0.04&        0.06&        0.04&        0.04\\
                    &      (0.19)&      (0.12)&      (0.15)&      (0.22)&      (0.24)&      (0.19)&      (0.24)&      (0.21)&      (0.19)\\
Sale details unknown (1=yes)&        0.18&        0.04&        0.04&        0.03&        0.02&        0.11&        0.02&        0.00&        0.01\\
                    &      (0.39)&      (0.19)&      (0.19)&      (0.17)&      (0.13)&      (0.31)&      (0.15)&      (0.04)&      (0.10)\\
\vspace{0.1em} \\ \emph{\textbf{B.Sales location}}&            &            &            &            &            &            &            &            &            \\
Sold within the village (1=yes)&        0.42&        0.50&        0.44&        0.72&        0.73&        0.66&        0.76&        0.77&        0.80\\
                    &      (0.49)&      (0.50)&      (0.50)&      (0.45)&      (0.44)&      (0.47)&      (0.43)&      (0.42)&      (0.40)\\
Sold within the district (1=yes)&        0.38&        0.48&        0.49&        0.28&        0.31&        0.25&        0.15&        0.16&        0.13\\
                    &      (0.49)&      (0.50)&      (0.50)&      (0.45)&      (0.46)&      (0.43)&      (0.35)&      (0.36)&      (0.34)\\
Sold outside the district (1=yes)&        0.05&        0.00&        0.06&        0.03&        0.06&        0.02&        0.15&        0.14&        0.13\\
                    &      (0.21)&      (0.07)&      (0.23)&      (0.18)&      (0.24)&      (0.14)&      (0.36)&      (0.35)&      (0.34)\\
Sale details unknown (1=yes)&        0.18&        0.04&        0.04&        0.03&        0.02&        0.11&        0.02&        0.00&        0.01\\
                    &      (0.39)&      (0.19)&      (0.19)&      (0.17)&      (0.13)&      (0.31)&      (0.15)&      (0.04)&      (0.10)\\
\midrule
Observations        &         996&       1,123&       1,007&       1,307&       1,219&       1,603&         986&       1,023&         984\\
\bottomrule

\end{tabular}
\begin{tablenotes}[para]
\footnotesize
\item Notes: Point estimates are sample means. Standard errors are in parentheses. 
\end{tablenotes}
\end{threeparttable}
\end{sidewaystable}

\begin{table}[htbp] 
\centering
\caption{Other variables by country and wave}
\label{tableA2}
\resizebox{\textwidth}{!}{%
\begin{threeparttable}
\begin{tabular}{@{\extracolsep{5pt}}l*{9}{c}}
\toprule
                    &\multicolumn{3}{c}{Ethiopia}          &\multicolumn{3}{c}{Nigeria}           &\multicolumn{3}{c}{Tanzania}          \\\cmidrule(lr){2-4}\cmidrule(lr){5-7}\cmidrule(lr){8-10}
                    & Wave 1 & Wave 2 & Wave 3 & Wave 1 & Wave 2 & Wave 3 & Wave 1 & Wave 2 & Wave 3 \\
\midrule
\multicolumn{10}{l}{\emph{\textbf{A. Agricultural characteristics}}} \\
\addlinespace
Total land size (hectares)&        1.60&        1.79&        1.73&        1.06&        0.92&        0.93&        2.43&        2.31&        2.38\\
                    &      (3.30)&      (4.56)&      (9.44)&      (1.71)&      (1.40)&      (1.38)&      (7.35)&      (4.01)&      (5.78)\\
Access to irrigation (1=yes)&        0.12&        0.11&        0.11&        0.04&        0.02&        0.02&        0.04&        0.04&        0.03\\
                    &      (0.33)&      (0.32)&      (0.32)&      (0.20)&      (0.15)&      (0.14)&      (0.20)&      (0.19)&      (0.18)\\
Access to fertilizer (1=yes)&        0.76&        0.81&        0.81&        0.50&        0.48&        0.59&        0.21&        0.31&        0.32\\
                    &      (0.43)&      (0.39)&      (0.39)&      (0.50)&      (0.50)&      (0.49)&      (0.41)&      (0.46)&      (0.47)\\
Experienced weather shocks (1=yes)&        0.20&        0.11&        0.32&        0.10&        0.16&        0.08&        0.48&        0.43&        0.41\\
                    &      (0.40)&      (0.31)&      (0.47)&      (0.30)&      (0.36)&      (0.28)&      (0.50)&      (0.50)&      (0.49)\\
\addlinespace
\multicolumn{10}{l}{\emph{\textbf{B. Household characteristics}}} \\
\addlinespace
Household size      &        5.41&        6.09&        6.59&        6.47&        6.84&        7.67&        5.56&        5.92&        5.85\\
                    &      (2.12)&      (2.22)&      (2.35)&      (3.13)&      (3.20)&      (3.50)&      (2.95)&      (3.20)&      (3.20)\\
Dependency ratio    &        1.47&        1.46&        1.98&        1.40&        1.62&        1.96&        1.74&        1.70&        1.77\\
                    &      (1.92)&      (1.67)&      (2.33)&      (1.94)&      (2.40)&      (2.67)&      (2.79)&      (2.77)&      (3.02)\\
Rural (1=yes)       &        0.98&        0.98&        0.98&        0.91&        0.91&        0.91&        0.90&        0.88&        0.87\\
                    &      (0.15)&      (0.15)&      (0.15)&      (0.29)&      (0.29)&      (0.29)&      (0.30)&      (0.33)&      (0.34)\\
Total consumption p.c. (real, LCU)&     4765.47&     4208.06&     3874.08&    88114.27&    73101.64&    49954.81&   529343.83&   463526.20&   512509.31\\
                    &   (4988.95)&   (2994.76)&   (2894.37)&  (67252.50)& (113158.61)&  (46084.16)& (367307.73)& (311567.92)& (347582.73)\\
Total consumption p.c. (real, USD)&      252.57&      223.03&      205.33&      559.53&      464.20&      317.21&      330.84&      289.70&      320.32\\
                    &    (264.41)&    (158.72)&    (153.40)&    (427.05)&    (718.56)&    (292.63)&    (229.57)&    (194.73)&    (217.24)\\
Experienced any shocks (1=yes)&        0.48&        0.34&        0.62&        0.34&        0.43&        0.35&        0.92&        0.84&        0.78\\
                    &      (0.50)&      (0.47)&      (0.49)&      (0.47)&      (0.50)&      (0.48)&      (0.28)&      (0.37)&      (0.41)\\
Access to a loan (1=yes)&        0.27&        0.31&        0.25&        0.22&        0.23&        0.19&        0.06&        0.08&        0.09\\
                    &      (0.44)&      (0.46)&      (0.43)&      (0.42)&      (0.42)&      (0.39)&      (0.23)&      (0.27)&      (0.29)\\
Consumption from own production&        0.74&        0.74&        0.76&        0.13&        0.13&        0.11&        0.26&        0.25&        0.20\\
                    &      (0.29)&      (0.28)&      (0.27)&      (0.30)&      (0.29)&      (0.24)&      (0.40)&      (0.40)&      (0.36)\\
Distance to nearest market (km)&       66.16&       66.20&       66.07&       73.24&       73.16&       73.15&       85.30&       84.92&       84.72\\
                    &     (47.20)&     (47.31)&     (47.04)&     (38.95)&     (38.91)&     (38.98)&     (54.15)&     (54.98)&     (54.95)\\
Access to transportation (1=yes)&        0.29&        0.30&        0.29&        0.23&        0.24&        0.42&        0.24&        0.22&        0.20\\
                    &      (0.45)&      (0.46)&      (0.45)&      (0.42)&      (0.43)&      (0.49)&      (0.43)&      (0.42)&      (0.40)\\
Transportation cost for sales (LCU)&        9.92&        5.42&       13.09&      906.94&      588.82&     1066.72&     1599.41&     3098.48&     2446.46\\
                    &     (61.95)&     (36.06)&    (171.21)&   (8030.09)&   (2638.48)&   (5810.97)&  (13845.92)&  (36199.91)&  (20471.87)\\
Transportation cost for sales (USD)&        0.53&        0.29&        0.69&        5.76&        3.74&        6.77&        1.00&        1.94&        1.53\\
                    &      (3.28)&      (1.91)&      (9.07)&     (50.99)&     (16.75)&     (36.90)&      (8.65)&     (22.62)&     (12.79)\\
\addlinespace
\multicolumn{10}{l}{\emph{\textbf{C. Household head's characteristics}}} \\
\addlinespace
Age of household head&       44.36&       46.03&       47.94&       49.91&       52.14&       53.58&       47.84&       49.84&       51.37\\
                    &     (14.70)&     (14.31)&     (14.41)&     (14.58)&     (14.64)&     (14.24)&     (15.34)&     (15.27)&     (15.10)\\
Education: Never attended school&        0.68&        0.68&        0.67&        0.51&        0.46&        0.43&        0.28&        0.30&        0.27\\
                    &      (0.47)&      (0.46)&      (0.47)&      (0.50)&      (0.50)&      (0.50)&      (0.45)&      (0.46)&      (0.45)\\
Education: Completed primary&        0.29&        0.29&        0.30&        0.21&        0.22&        0.26&        0.65&        0.62&        0.64\\
                    &      (0.45)&      (0.45)&      (0.46)&      (0.41)&      (0.41)&      (0.44)&      (0.48)&      (0.48)&      (0.48)\\
Education: Completed secondary+&        0.03&        0.03&        0.03&        0.28&        0.32&        0.31&        0.08&        0.07&        0.07\\
                    &      (0.16)&      (0.16)&      (0.18)&      (0.45)&      (0.47)&      (0.46)&      (0.26)&      (0.25)&      (0.25)\\
Married household head (1=yes)&        0.85&        0.82&        0.81&        0.86&        0.86&        0.83&        0.78&        0.78&        0.75\\
                    &      (0.36)&      (0.38)&      (0.39)&      (0.34)&      (0.35)&      (0.38)&      (0.42)&      (0.41)&      (0.43)\\
\midrule
Observations        &        2118&        2118&        2118&        2189&        2189&        2189&        1704&        1704&        1704\\
\bottomrule
\end{tabular}
\begin{tablenotes}
\item \footnotesize Notes: Point estimates are sample means. Standard errors are in parentheses. For Ethiopia, "consumption from own production" is a self-reported variable. For Nigeria and Tanzania, this variable was calculated based on the consumption module questions. All monetary values are converted from local currency units (LCU) to constant 2013 U.S. dollars to control for inflation. The conversion uses the 2013 period-average exchange rates from the World Bank: 1 USD = 18.87 Ethiopian Birr, 1 USD = 1,600 Tanzanian Shilling, and 1 USD = 157.48 Nigerian Naira.
\end{tablenotes}
\end{threeparttable}
} 
\end{table}

\begin{table}[htbp]\centering
\begin{threeparttable}
\def\sym#1{\ifmmode^{#1}\else\(^{#1}\)\fi}
\caption{Effects of gender on market types}
\label{tableA3}
\begin{tabular}{@{\extracolsep{5pt}}l*{3}{c}}
\toprule
                    & \multicolumn{1}{c}{(1)} & \multicolumn{1}{c}{(2)} & \multicolumn{1}{c}{(3)} \\
                    & \multicolumn{1}{c}{Sold informal} & \multicolumn{1}{c}{Sold formal} & \multicolumn{1}{c}{Sold others} \\
\midrule
\multicolumn{4}{l}{\textbf{Panel A: Ethiopia}} \\
Woman household head (1=yes)&      -0.004         &      -0.011         &      -0.020\sym{*}  \\
                    &     (0.041)         &     (0.039)         &     (0.012)         \\
Observations        &        2977         &        2977         &        2977         \\
\addlinespace
Share of women-managed  land area&       0.004         &      -0.014         &      -0.013         \\
                    &     (0.038)         &     (0.037)         &     (0.011)         \\
Observations        &        2973         &        2973         &        2973         \\
\midrule
\multicolumn{4}{l}{\textbf{Panel B: Nigeria}} \\
Woman household head (1=yes)&       0.105\sym{***}&      -0.027         &      -0.020         \\
                    &     (0.040)         &     (0.034)         &     (0.022)         \\
Observations        &        4045         &        4045         &        4045         \\
\addlinespace
Share of women-managed  land area&       0.044\sym{*}  &       0.034         &      -0.024\sym{*}  \\
                    &     (0.026)         &     (0.026)         &     (0.014)         \\
Observations        &        4039         &        4039         &        4039         \\
\midrule
\multicolumn{4}{l}{\textbf{Panel C: Tanzania}} \\
Woman household head (1=yes)&      -0.006         &       0.003         &      -0.008         \\
                    &     (0.031)         &     (0.034)         &     (0.011)         \\
Observations        &        2945         &        2945         &        2945         \\
\addlinespace
Share of women-managed  land area&       0.022         &      -0.020         &      -0.004         \\
                    &     (0.031)         &     (0.032)         &     (0.011)         \\
Observations        &        2833         &        2833         &        2833         \\
\bottomrule
\end{tabular}
\begin{tablenotes}[para]
\footnotesize
\item Notes: Standard errors clustered at the household level are in parentheses. All regressions include a standard set of control variables for household demographics (size, dependency ratio, head’s age, marital status, and education), agricultural and economic characteristics (land size, irrigation, loan access, and distance to market), and exposure to climatic shocks. Level of statistical significance: \sym{*} \(p<0.10\), \sym{**} \(p<0.05\), \sym{***} \(p<0.01\).
\end{tablenotes}
\end{threeparttable}

\end{table}

\begin{table}[htbp]\centering

\begin{threeparttable}
\caption{Effects of gender on sales locations}
\label{tableA4}
\begin{tabular}{@{\extracolsep{5pt}}l*{3}{c}}
\toprule
                    &      (1)        &       (2)        &         (3)           \\
                    & Sold in Village & Sold in District & Sold Outside District \\
\midrule
\multicolumn{4}{l}{\textbf{Panel A: Ethiopia}} \\
Woman household head (1=yes) & -0.070      & 0.036       & -0.005      \\
                              & (0.043)     & (0.043)     & (0.012)     \\
Observations                  & 2977        & 2977        & 2977        \\
\addlinespace
Share of women-managed  land area & -0.014      & -0.019      & 0.001       \\
                                  & (0.040)     & (0.040)     & (0.012)     \\
Observations                      & 2973        & 2973        & 2973        \\
\midrule
\multicolumn{4}{l}{\textbf{Panel B: Nigeria}} \\
Woman household head (1=yes) & 0.056       & 0.017       & -0.010      \\
                              & (0.046)     & (0.040)     & (0.011)     \\
Observations                  & 4045        & 4045        & 4045        \\
\addlinespace
Share of women-managed  land area & 0.067\sym{**} & -0.001      & -0.017\sym{***} \\
                                  & (0.031)       & (0.029)     & (0.006)       \\
Observations                      & 4039          & 4039        & 4039          \\
\midrule
\multicolumn{4}{l}{\textbf{Panel C: Tanzania}} \\
Woman household head (1=yes) & 0.005       & -0.013      & -0.011      \\
                              & (0.031)     & (0.025)     & (0.027)     \\
Observations                  & 2945        & 2945        & 2945        \\
\addlinespace
Share of women-managed  land area & 0.009       & -0.004      & -0.020      \\
                                  & (0.029)     & (0.024)     & (0.023)     \\
Observations                      & 2833        & 2833        & 2833        \\
\bottomrule
\end{tabular}
\begin{tablenotes}[para]
\footnotesize
\item Notes: Standard errors clustered at the household level are in parentheses. All regressions include a standard set of control variables for household demographics (size, dependency ratio, head’s age, marital status, and education), agricultural and economic characteristics (land size, irrigation, loan access, and distance to market), and exposure to climatic shocks. Level of statistical significance: \sym{*} \(p<0.10\), \sym{**} \(p<0.05\), \sym{***} \(p<0.01\).
\end{tablenotes}
\end{threeparttable}
\end{table}

\begin{table}[htbp]
\centering
\caption{Effects of gender on commercialization: Pooled OLS results}
\label{tableA5}
\begin{threeparttable}

\begin{tabular}{@{\extracolsep{5pt}}l*{3}{c}}
\toprule
                    &\multicolumn{1}{c}{(1)}&\multicolumn{1}{c}{(2)}&\multicolumn{1}{c}{(3)}\\
                    &\multicolumn{1}{c}{Sales (HH)}&\multicolumn{1}{c}{CCI}&\multicolumn{1}{c}{Cash Crop Ratio}\\
\midrule
\multicolumn{4}{l}{\textbf{Panel A: Ethiopia}} \\
Woman household head (1=yes)&     -0.082\sym{***}&      -0.038\sym{***}&      -0.087\sym{***}\\
                    &     (0.027)         &     (0.012)         &     (0.021)         \\
\addlinespace
Observations        &        6072         &        6072         &        6072         \\
\addlinespace
Share of women-managed  land area&      -0.097\sym{***}&      -0.036\sym{***}&      -0.080\sym{***}\\
                    &     (0.025)         &     (0.011)         &     (0.020)         \\
\addlinespace
Observations        &        6065         &        6065         &        6065         \\
\midrule
\multicolumn{4}{l}{\textbf{Panel B: Nigeria}} \\
Woman household head (1=yes)&       -0.077\sym{**} &      -0.061\sym{**} &      -0.066\sym{*}  \\
                    &     (0.036)         &     (0.027)         &     (0.035)         \\
\addlinespace

Observations        &        6394         &        6394         &        6394         \\
\addlinespace
Share of women-managed  land area&       -0.089\sym{***}&      -0.044\sym{**} &      -0.103\sym{***}\\
                    &     (0.025)         &     (0.018)         &     (0.024)         \\
\addlinespace
Observations        &        6387         &        6387         &        6387         \\
\midrule

\multicolumn{4}{l}{\textbf{Panel C: Tanzania}} \\
Woman household head (1=yes)&      -0.030         &      -0.006         &      -0.029         \\
                    &     (0.028)         &     (0.018)         &     (0.028)         \\
\addlinespace
Observations        &        5003         &        5003         &        5003         \\
\addlinespace
\addlinespace
Share of women-managed  land area&       -0.041         &      -0.013         &      -0.046\sym{*}  \\
                    &     (0.026)         &     (0.015)         &     (0.026)         \\
\addlinespace
Observations        &        4750         &        4750         &        4750         \\
\bottomrule
\end{tabular}
\begin{tablenotes}[para]
\footnotesize
\item Notes: Standard errors clustered at the household level are in parentheses. All regressions include a standard set of control variables for household demographics (size, dependency ratio, head’s age, marital status, and education), agricultural and economic characteristics (land size, irrigation, loan access, and distance to market), and exposure to climatic shocks. CCI represents the crop commercialization index, calculated as the ratio of the gross value of crops sold to the gross value of crops produced per season or year.  The cash crop ratio is calculated as the proportion of output sold that comes from cash crops. Level of statistical significance: \sym{*} \(p<0.10\), \sym{**} \(p<0.05\), \sym{***} \(p<0.01\).
\end{tablenotes}
\end{threeparttable}
\end{table}

\end{document}